\documentclass[sigconf,nonacm]{acmart}
\usepackage{multirow}
\usepackage{subfigure}
\usepackage{blindtext}
\usepackage{enumitem}
\usepackage{algorithm}  
\usepackage{algpseudocode}  
\usepackage{amsmath}  
\usepackage{hyperref}
\usepackage{flushend}
\usepackage{balance}

\AtBeginDocument{%
\providecommand\BibTeX{{%
  \normalfont B\kern-0.5em{\scshape i\kern-0.25em b}\kern-0.8em\TeX}}}

  \copyrightyear{2022} 
  \acmYear{2022} 
  \setcopyright{acmcopyright}
  \acmConference[CIKM '22]{Proceedings of the 31st ACM International Conference on Information and Knowledge Management}{October 17--21, 2022}{Atlanta, GA, USA.}
  \acmBooktitle{Proceedings of the 31st ACM International Conference on Information and Knowledge Management (CIKM '22), October 17--21, 2022, Atlanta, GA, USA}
  \acmPrice{15.00}
  \acmDOI{10.1145/xxxxxx.xxxxxx}
  \acmISBN{978-1-4503-9236-5/22/10}
\begin{document}

\title{Hybrid Transfer in Deep Reinforcement Learning for Ads Allocation}



\author{Ze Wang}
\authornote{Equal contribution. Listing order is random.}
\affiliation{%
 \institution{Meituan}
 \city{Beijing}
 \country{China}
}
\email{wangze18@meituan.com}

\author{Guogang Liao}
\authornotemark[1]
\affiliation{%
 \institution{Meituan}
 \city{Beijing}
 \country{China}
}
\email{liaoguogang@meituan.com}

\author{Xiaowen Shi}
\authornote{Corresponding author.}
\affiliation{%
 \institution{Meituan}
 \city{Beijing}
 \country{China}
}
\email{shixiaowen03@meituan.com}

\author{Xiaoxu Wu}
\affiliation{%
 \institution{Meituan}
 \city{Beijing}
 \country{China}
}
\email{wuxiaoxu04@meituan.com}

\author{Chuheng Zhang}
\authornote{This work was done when Chuheng Zhang was an intern in Meituan.}
\affiliation{%
 \institution{Meituan}
 \city{Beijing}
 \country{China}
}
\email{zhangchuheng123@live.com}

\author{Bingqi Zhu}
\affiliation{%
 \institution{Meituan}
 \city{Beijing}
 \country{China}
}
\email{zhubingqi@meituan.com}

\author{Yongkang Wang}
\affiliation{%
 \institution{Meituan}
 \city{Beijing}
 \country{China}
}
\email{wangyongkang03@meituan.com}

\author{Xingxing Wang}
\affiliation{%
 \institution{Meituan}
 \city{Beijing}
 \country{China}
}
\email{wangxingxing04@meituan.com}

\author{Dong Wang}
\affiliation{%
 \institution{Meituan}
 \city{Beijing}
 \country{China}
}
\email{wangdong07@meituan.com}

\renewcommand{\shortauthors}{Ze Wang and Guogang Liao, al.}

\begin{abstract}
  Ads allocation, which involves allocating ads and organic items to limited slots in feed with the purpose of maximizing platform revenue, has become a research hotspot. Notice that, platforms (e.g., e-commerce platforms, video platforms, food delivery platforms and so on) usually have multiple entrances for different categories and some entrances have few visits. Data from these entrances has low coverage, which makes it difficult for the agent to learn. To address this challenge, we propose \textbf{S}imilarity-based \textbf{H}ybrid  \textbf{T}ransfer for \textbf{A}ds  \textbf{A}llocation (SHTAA), which effectively transfers samples as well as knowledge from data-rich entrance to  data-poor entrance. Specifically, we define an uncertainty-aware similarity for MDP to estimate the similarity of MDP for different entrances. Based on this similarity, we design a hybrid transfer method, including instance transfer and strategy transfer, to efficiently transfer samples and knowledge from one entrance to another. Both offline and online experiments on Meituan food delivery platform demonstrate that the proposed method could achieve better performance for data-poor entrance and increase the revenue for the platform.
\end{abstract}


\maketitle

\section{Introduction}
\begin{figure}[tb]
  \centering
  \includegraphics[width=1\linewidth,height=0.48\textwidth]{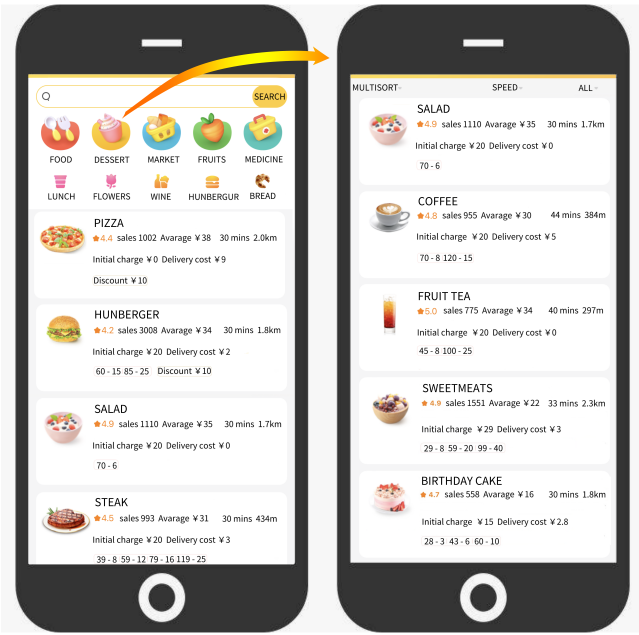}
  \caption{
    The Meituan food delivery platform has multiple entrances for categories (e.g., Homepage, Desert and so on), facilitating users to access different categories of content.
  }
  \label{fig:fig1}
\end{figure}

Ads and organic items are mixed together and displayed to users in e-commerce feed nowadays \cite{yan2020LinkedInGEA,Ghose2009AnEA,li2020deep}. How to allocate limited ads slots reasonably and effectively to maximize platform revenue has attracted growing attention \cite{Wang2011LearningTA,Mehta2013OnlineMA,zhang2018whole}. Several recent strategies for ads allocation model the problem as Markov Decision Process (MDP) \cite{sutton1998introduction} and solve it using reinforcement learning (RL) \cite{zhang2018whole,liao2021cross,zhao2019deep, Feng2018LearningTC, zhao2020jointly}. For instance, 
\citet{xie2021hierarchical} propose a hierarchical RL-based framework to first decide the type of the item to present and then determine the specific item for each slot.
\citet{liao2021cross} propose CrossDQN which takes the crossed items according to the action as input and allocates the slots in one screen at a time.

However, these excellent RL-based algorithms face one major challenge when applied in platforms (e.g., e-commerce platforms, video platforms, food delivery platforms and so on). As shown in Figure \ref{fig:fig1}, the Meituan food delivery platform has multiple entrances for different categories (e.g., Homepage, Food, Desert and so on), facilitating users to access different categories of content. Some entrances have few visits, resulting in a low data coverage. This further makes it difficult for the agent to learn.
Therefore, it is desirable for a learning algorithm to transfer samples and leverage knowledge acquired in data-rich entrance to improve performance of the agent for ads allocation in data-poor entrance. 
Motivated by this phenomenon, we incorporate Transfer learning (TL) technology into RL to solve this problem.

TL in RL has been rarely used for ads allocation. Meanwhile, TL in RL for other scenarios \cite{zhu2020transfer,giannopoulos2021deep,tao2021repaint, liu2019value, tirinzoni2018importance}. For instance, \citet{tirinzoni2018importance} present a algorithm called IWFQI for transferring samples in batch RL that uses importance weighting to automatically account for the difference in source and target distributions. But IWFQI does not fully exploit possible similarities between tasks. \citet{liu2019value} quantify the environmental dynamics of an MDP by the N-step return (NSR) values and present a knowledge transfer method called NSR-based value function transfer. However, they ignore the uncertainty of the NSR model itself when transferring, since the rewards can induce randomness in observed long-term return \cite{dabney2018distributional}. 

To this end, we present Similarity-based Hybrid Transfer for Ads Allocation (SHTAA)\footnote{The code and data example are publicly accessible at \url{https://github.com/princewen/SHTAA}}, which effectively transfers samples and knowledge from data-rich entrance to agent trained from other data-poor entrance. Specifically, we quantify the {environmental} dynamics of an MDP using distributional N-step return (NSR) value, which is predicted by pre-trained NSR model. And we define a concept of uncertainty-aware similarity of MDP based on it.
Then we design a hybrid transfer framework,  which consists of instance transfer and strategy transfer, based on the similarity value. The framework aims to effectively transfer the samples and  knowledge from data-rich source task to data-poor target task.
We conduct extensive offline experiments and {evaluate} our approach on Meituan food delivery platform. The experimental results show that {SHTAA} can effectively improve the performance of {the agent for ads allocation in data-poor entrance,} and obtain significant improvements in platform revenue.

\section{Problem Formulation}
\label{sec:problem}

On Meituan food delivery platform, we present $K$ slots in one screen and allocate for each screen sequentially. So the ads allocation problem for different entrances can be formulated as a MDP, using a tuple ($\mathcal{S}$, $\mathcal{A}$, $r$, ${P}$, $\gamma$), which specifies the state space $\mathcal{S}$, the action space $\mathcal{A}$, the reward $r$, the state transition probability $P$ and the discount factor $\gamma$. Since this paper mainly focuses on how to transfer knowledge from data-rich entrance to data-proo entrance to improve the performance of the agent for ads allocation, we mainly follow the recent work - CrossDQN \cite{liao2021cross} in the detailed definition of elements (i.e., $\mathcal{S}$, $\mathcal{A}$, $r$, ${P}$, $\gamma$) on ads allocation, to simplify our content and reduce the cost of understanding for readers.
Besides, we define the N-step return (NSR) \cite{liu2019value} after current state $s_t$ as follows: 
\begin{equation}
  \begin{aligned}
    r^N = r_t + \gamma \cdot r_{t+1} + \dots + \gamma^{N-1}\! \cdot r_{t+N-1} 
  \end{aligned}
  \label{eq:eta}
\end{equation}

We denote the dataset for source task as $\mathcal{D}_{S}=\{(s,a,r,r^N,s')\}$, and the dataset for target task as $\mathcal{D}_{T}=\{(s,a,r,r^N,s')\}$. The objective is to find an ads allocation policy for {the} target task {based} on $\mathcal{D}_{S}$ and $\mathcal{D}_{T}$ to maximize {the cumulative reward}.

\section{Methodology}
In this section, we {will} introduce SHTAA in detail. 
Two main ideas {are}: i) using predicted uncertain-aware NSR to measure MDP similarity, ii) proposing a hybrid transfer approach (consisting of instance transfer and strategy transfer) {for selective transfer and avoidance of negative transfer}.
\subsection{Uncertainty-Aware MDP Similarity}
\citet{liu2019value} demonstrate the effectiveness of using NSR to quantify the environmental dynamics of an MDP and transfer based on it. But they ignore the uncertainty of the NSR model itself. In this paper, we propose an uncertainty-aware MDP similarity concept in which we combine high-dimensional representation capability of deep learning model and distribution modeling capability of Gaussian process (GP) \cite{rasmussen2003gaussian} to construct the NSR model. 

Specifically, we assume a GP prior distribution over $f$, i.e., $f \sim \mathcal{GP}(m(s,a), k((s,a),(s,a)'))$, where $m(s,a) = E[f(s,a)]$ is the mean function and $k((s,a),(s,a)')$ is the covariance kernel function. Formulately, we define $f: (\mathcal{S},\mathcal{A}) \rightarrow \mathbb{R}$ as the prediction model of NSR. Under a Bayesian perspective of function learning, the uncertainty of NSR can be measured by the uncertainty of function $f$. We impose a prior distribution $p(f)$ to the function $f$ and (approximately) infer the posterior distribution $p(f|\mathcal{D})$ over functions as the learned model:
\begin{equation}
  \begin{aligned}
    p(f|\mathcal{D}) = \frac{p(f)p(\mathcal{D}|f)}{p(\mathcal{D})}  
  \end{aligned}
\end{equation}

We aim to estimate the function $f^*=f\big((s,a)_*\big)$ as well as its uncertainty on a test sample $(s,a)_*$, given the observed samples $\mathcal{D}^b = \Big\{\big((s_{i},a_{i}), r^N\big)\Big\}_{i=1}^{b}$ at batch $B$. 
Let $(\mathbf{S},\mathbf{A})_b \in\mathbb{R}^{b\times d}$ denote the matrix of the observed feature vectors, $\mathbf{f}_{b}=f\big((\mathbf{S},\mathbf{A})_b\big)\in\mathbb{R}^{b}$ be the vector of the corresponding function values, and $\mathbf{y}_{b}=[r^{N}_{1},\cdots,r^{N}_{b}]^\top\in\{0,1\}^{b}$ denote the vector of user feedback. Then the likelihood of $r^N, \mathbf{f}_{b}$ and $f_*$ is:
\begin{equation}
  p(\mathbf{y}_{b},\mathbf{f}_{b},f_*)=p(\mathbf{y}_{b}|\mathbf{f}_{b})p(\mathbf{f}_{b},f_*)=p(\mathbf{f}_{b},f_*)\prod{}_{i=1}^b p(y_{i}|f_{i}),
\end{equation}
where $p(\mathbf{f}_{b},f_*)$ is a multivariate Gaussian distribution defined by the GP prior:

\begin{equation}
  \!p(\mathbf{f}_{b},f^*)\!=\!\mathcal{N}\!\left(\!\!
    \begin{array}{c|}
      \! \mathbf{f}_{t}\!\!\! \\
      \! f^*\!\!\!
    \end{array}
    \!
    \begin{bmatrix}
      
      m\big((\mathbf{S},\!\mathbf{A})_b\big) \\
      m\big((s,\!a)_*\big)
      \!
    \end{bmatrix}
    \!
    ,
    \!
    \begin{bmatrix}
        k\big((\mathbf{S},\!\mathbf{A})_{b},\!(\mathbf{S},\!\mathbf{A})_b\big) \!\!\!&\!\!k\big(\mathbf{S},\!\mathbf{A})_b,\!(s,\!a)_*\big)\\
        k\big((s,\!a)_*,\!(\mathbf{S},\!\mathbf{A})_b) \!\!\!& \!\!k\big((s,\!a)_*,\!(s,\!a)_*\big)\\
    \end{bmatrix}\right),
\end{equation}

The posterior distribution of $f_*$ given $\mathcal{D}^b$ can be evaluated as:
\begin{equation}\label{eqn:exact-inf}
  \!\!p(f_*|\mathbf{y}_{b})\!=\!\frac{\int p(\mathbf{y}_{b},\mathbf{f}_{b},f_*) \mathrm{d}\mathbf{f}_{b}}{p(\mathbf{y}_{b})}\!=\!\frac{1}{p(\mathbf{y}_{b})}\int p(\mathbf{y}_{b}|\mathbf{f}_{b})p(\mathbf{f}_{b},f^*) \mathrm{d}\mathbf{f}_{b},
\end{equation}
which is the distributional prediction we desire. According to the solving skills for GP in high-dimensional sample space \cite{salimbeni2017doubly, du2021exploration}, we can calculate the predicted mean $\hat \mu$ and variance $\hat \Sigma$ of NSR. We denote the predicted distribution of NSR as follows:
\begin{equation}
  \begin{aligned}
    p_S(s,a) = \mathcal{N} (\hat \mu_S, \hat \Sigma_S) \\ 
    p_T(s,a) = \mathcal{N} (\hat \mu_T, \hat \Sigma_T) \\ 
  \end{aligned}
\end{equation}
where $p_S(s,a)$ and $p_T(s,a)$ are the predicted distributions from NSR models pre-trained with source and target dataset, respectively.

Subsequently, we {calculate} the uncertainty-aware MDP similarity based on the Jensen–Shannon divergence \cite{menendez1997jensen} between $p_S(s,a)$ and $p_T(s,a)$. The similarity-based weight $w$ for a sample is calculated as follows:
\begin{equation}
  \begin{aligned}
    w = 1-\text{JS}\big(p_S(s,a) || p_T(s,a)\big) 
  \end{aligned}
  \label{eq:sim}
\end{equation}

\begin{figure}[tb]
  \centering
  \includegraphics[width=1\linewidth]{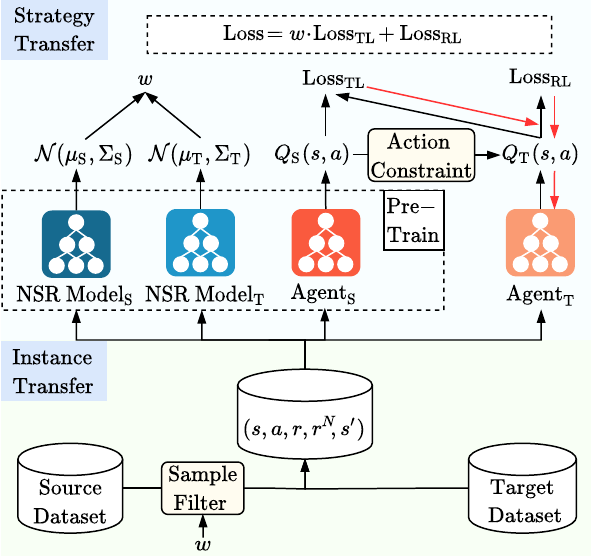}
  \caption{
  The hybrid transfer includes instance transfer and strategy transfer. {Red lines indicate gradient propagation}.
  }
  \label{fig:fig2}
\end{figure}

\subsection{Hybrid Transfer}
The hybrid transfer method, which is shown in Figure \ref{fig:fig2}, consists of instance transfer and strategy transfer. Next we will introduce each part {separately}.
\subsubsection{Instance Transfer}
The key challenge in instance transfer is to avoid the negative transfer. In this paper, we propose an instance transfer method based on the uncertainty-aware MDP similarity. Specifically, given a sample $(s,a,r,r^N\!\!,s',w) \in \mathcal{D}_S$ and a weight threshold $\tau$, the local environmental dynamics of $(s,a)$ in the source task is regarded similar to that in the target task if $w \geq \tau$. In this case, this sample for the source task can be added directly into the mixed dataset. 
 And when $w < \tau$, which means that the local environmental dynamics related to $(s,a)$ in the source task and target task are different, this sample will be filtered out.\footnote{We have tried weighted transfer in instance transfer, the performances are close.} In this way, the selective instance transfer based on uncertainty-aware MDP similarity can effectively avoid negative transfer.

\subsubsection{Strategy Transfer}
The pre-trained agent for the source task (hereinafter referred to as $\text{agent}_S$) can guide the learning of the agent for the target task (hereinafter referred to as $\text{agent}_T$). Specifically, we first constrain the action space of $\text{agent}_T$'s target value function
based on the output Q-value of $\text{agent}_S$. The RL loss for $\text{agent}_T$ is calculated as follows:
\begin{equation}
  \begin{aligned}
    \text{Loss}_{\text{RL}}=\big( r + \gamma \max_{a'\in\mathcal{A_T}} Q_T(s', a') - Q_T(s,a) \big) ^ 2,
  \end{aligned}
\end{equation}
where $\mathcal{A_T}$ is a set of $\beta$ actions corresponding to the top-$\beta$ highest $Q_S(s', a_i)$ and $\beta$ is determined by $w$. By restricting the action space, the $\text{agent}_T$ are forced {to behave similar to $\text{agent}_S$}. 

{Second}, the output Q-value of $\text{agent}_S$ can also {guide} the learning process of $\text{agent}_T$. We take $Q_S(s,a)$ as the learning target of $\text{agent}_T$:
\begin{equation}
  \begin{aligned}
    \text{Loss}_{\text{TL}} = \big(Q_S{(s,a)}-Q_T(s,a)\big)^2.
  \end{aligned}
\end{equation}

The similarity-based weight is used to adjust the extent of strategy transfer, w.r.t. the loss:

\begin{equation}
  \label{eq:loss}
  \begin{aligned}
    \text{Loss} =\  \frac{1}{|B|}\sum_{(s,a,r,r^N\!,s',w)\in B} \Big( w\cdot\text{Loss}_{\text{TL}}  + \text{Loss}_{\text{RL}} \Big).
  \end{aligned}
\end{equation}

\subsection{Offline Training}
We follows the offline RL paradigm, and the process of offline training is shown in Algorithm \ref{alg:offline}.
We first pre-train the NSR models and $\text{agent}_S$. Then we train the $\text{agent}_T$ through SHTAA. 

\begin{algorithm}[hbp]  
  \renewcommand\arraystretch{1.1}
  \caption{Offline training of SHTAA}
  \label{alg:offline}
  \begin{algorithmic}[1] 
    \State Source dataset $\mathcal{D}_{S}$, target dataset $\mathcal{D}_{T}$
    \State $\textbf{pre-train}$
    \State $\ \ \ \ \ $Train the NSR Model for source task on $\mathcal{D}_{S}$ 
    \State $\ \ \ \ \ $Train the NSR Model for target task on $\mathcal{D}_{T}$ 
    \State $\ \ \ \ \ $Train the $\text{agent}_S$ on $\mathcal{D}_{S}$ 
    \State $\textbf{train}$
    
    \State $\ \ \ \ $ Calculate the similarity-based weight $w$ for each sample
    \State $\ \ \ \ $ Filter samples in $\mathcal{D}_S$ based on weight $w$ and threshold $\tau$
    \State $\ \ \ \ $ Merge filtered $\mathcal{D}_{S}$ and $\mathcal{D}_{T}$ as $\mathcal{D}=\{(s,a,r,r^N\!\!,s',w)\}$
    \State $\ \ \ \ \ $Initialize $\text{agent}_T$ with random weights
    \State $\ \ \ \ \textbf{repeat}$
        \State $\ \ \ \ \ \ \ \ $ Sample a batch $B$ of $(s,a,r,r^N\! \! ,s',w)$ from $\mathcal{D}$
        \State $\ \ \ \ \ \ \ \ $ Update network parameters by minimizing $\text{Loss}$ in \eqref{eq:loss}
     \State $\ \ \ \ \textbf{until}$ Convergence
  \end{algorithmic}  
  \end{algorithm}

\subsection{Online Serving} 
  In the online serving system, the agent for ads allocation in target entrance selects the action with the highest reward 
  based on current state and converts the action to ads slots set for the output. When user pulls down, state is updated and the above process is repeated.

\section{Experiments}
\subsection{Experimental Settings}
\subsubsection{Dataset}
We collect the dataset by running an exploratory policy on Meituan food delivery platform during January 2022. The dataset contains 12,411,532 requests from 1,919,201 users in the source entrance and 813,271 requests from 214,327 users in the target entrance. Each request contains several transitions.

\subsubsection{Evaluation Metrics}
We evaluate different methods with ads revenue $R^\text{ad}$ and service fee $R^\text{fee}$. Follow the definition in \cite{liao2021cross}.

\subsubsection{Parameters Settings}
The hidden layer sizes of the NSR models is $(128, 64, 32)$. The structure and parameters of the agents follow the work in CrossDQN \cite{liao2021cross}.
The learning rate is $10^{-3}$, the optimizer is Adam \cite{kingma2014adam} and the batch size is 8,192. $\tau$ is $0.7$ and $N$ is 3.

\subsection{Offline Experiment}
In this section, we validate our method on offline data and evaluate the performance using an offline estimator. 
Through extended engineering, the offline estimator models the user preference and aligns well with the online service.

\subsubsection{Baselines \& Ablations}
We study four baselines and three ablated variants to verifies the effectiveness of SHTAA. 
\begin{itemize}[leftmargin=*]
  \item \textbf{DEAR}  \cite{zhao2019deep} is an advanced DQN architecture to jointly determine three related tasks for ads allocation. Here we train it on $\mathcal{D}_T$.
  \item \textbf{CrossDQN} \cite{liao2021cross} is an advanced method for ads allocation. Here we take it as the structure of $\text{agent}_T$ and train $\text{agent}_T$ on $\mathcal{D}_T$.
  \item \textbf{Cross DQN (w/ $\mathcal{D}_S$)} transfers all samples in source dataset into the training of $\text{agent}_T$ based on the previous baseline.
  \item  \textbf{IWFQI} \cite{tirinzoni2018importance} is a  algorithm for transferring samples in batch RL that uses importance weighting to automatically account for the difference in the source and target distributions.
  \item  \textbf{NSR-CrossDQN}. \citet{liu2019value} propose the NSR-based value function transfer method.Here we implement this transfer method on $\mathcal{D}_S$ and $\mathcal{D}_T$ based on CrossDQN.
  \item \textbf{SHTAA (w/o UA-Sim)} does not use uncertainty-aware MDP similarity and uses the MDP similarity concept defined in the NSR-based value function transfer method \cite{liu2019value} instead.
  \item \textbf{SHTAA (w/o AC)} does not use action constraint in SHTAA.
  \item \textbf{SHTAA (w/o $\text{Loss}_\text{TL}$)} does not use $\text{Loss}_\text{TL}$ in SHTAA.
\end{itemize}

\subsubsection{Performance Comparison}
We present the experimental results in Table \ref{result} and have the following findings:
i) The performance of SHTAA is superior to all baselines, which mainly justifies that SHTAA can selectively transfer the samples and knowledge from source task to target task.
ii) Compared with IWFQI, the superior performance of our method mainly justifies the effectiveness of our strategy transfer.
iii) Compared with NSR-CrossDQN, the superior performance of our method mainly justifies the effectiveness of our uncertainty-aware MDP similarity concept and action constraint. 

\subsubsection{Ablation Study}
To verify the impact of our designs, we study three ablated variants of our method and have the following findings:
i) The performance of SHTAA is superior to CrossDQN, which verifies the effectiveness of all our designs.
ii) The performance of SHTAA is superior to SHTAA (w/o UA-Sim), which mainly verifies the effectiveness of uncertainty-aware MDP similarity. iii) The performance gap between SHTAA and SHTAA (w/o AC) indicates the effectiveness of action constraint. iv) The performance gap between SHTAA and SHTAA (w/o $\text{Loss}_\text{TL}$) indicates the guiding significance of $\text{agent}_S$.

\subsubsection{Hyperparameter Analysis}
We analyze the sensitivity of these two hyperparameters: $N$ and $\tau$. The optimal hyperparameter values are obtained by grid search and we have the following findings: 
i) $N$ is the step number of NSR. Due to the discount rate $\gamma$, the rewards sampled at larger time steps must contribute less to the NSR. Therefore, if an appropriate step number $N$ is chosen, the corresponding NSR may contain the most information about the local environmental dynamics. ii) $\tau$ is the weight threshold for instance transfer. A smaller $\tau$ would lead to more negative transfer while a larger $\tau$ would lead to less efficient transfer learning.

\subsection{Online Results}
We deploy the agent trained by SHTAA on Meituan food delivery platform's  data-poor entrance through online A/B test. The ads revenue and service fees increase by 3.72\% and 3.31\%, which demonstrates that our method greatly increases the platform revenue.

\begin{table}[!tb]
  \caption{The experimental results. Each result is presented in the form of mean $\pm$ standard deviation.}
  \centering
  \renewcommand\arraystretch{1.15}
  \setlength{\tabcolsep}{1.9mm}
  \begin{tabular*}{0.94\linewidth}{l|c|c}
  \hline
  \rule{0pt}{11.4pt}
      model & $R^{\text{ad}}$ & $R^{\text{fee}}$ \\ \hline  \hline
      DEAR & 0.2389\ ($\pm$0.0008) & 0.2471\ ($\pm$0.0013) \\ \hline  
      CrossDQN & 0.2450\ ($\pm$0.0004) & 0.2492\ ($\pm$0.0007) \\ \hline 
      CrossDQN (w/ $\mathcal{D}_S$) & 0.2461\ ($\pm$0.0005) & 0.2497\ ($\pm$0.0009) \\ \hline 
      IWFQI & 0.2463\ ($\pm$0.0004) & 0.2505\ ($\pm$0.0007) \\ \hline 
      NSR-CrossDQN & 0.2471\ ($\pm$0.0009) & 0.2518\ ($\pm$0.0009) \\ \hline 
      \textbf{SHTAA} & \textbf{0.2562\ ($\pm$0.0003)} & \textbf{0.2596\ ($\pm$0.0005)} \\ \hline
      \ - w/o UA-Sim & 0.2465\ ($\pm$0.0004) & 0.2521\ ($\pm$0.0001) \\ \hline
      \ - w/o AC & 0.2503\ ($\pm$0.0001) & 0.2561\ ($\pm$0.0002) \\ \hline
      \ - w/o $\text{Loss}_\text{TL}$ & 0.2527\ ($\pm$0.0006) & 0.2572\ ($\pm$0.0006) \\ \hline \hline 
      Improvement & 3.68\% & 3.09\% \\ \hline
  \end{tabular*}
  \label{result}
\end{table}

\section{Conclusions}
In this paper, we propose Similarity-based Hybrid Transfer for Ads Allocation (SHTAA) to effectively transfer the samples and knowledge from data-rich source entrance to other data-poor entrance. 
Both offline experiments and online A/B test have demonstrated the superior performance and efficiency of our method.

\balance
\begin{acks}
  We thank the anonymous reviewers for their suggestions and comments. We also thank Fan Yang, Hui Niu for helpful discussions.
\end{acks}
\bibliographystyle{ACM-Reference-Format}
\bibliography{transfer}

\appendix

\end{document}